\documentclass[%
 reprint,
 superscriptaddress,
 longbibliography,
 amsmath,amssymb,
 aps,
]{revtex4-2}

\usepackage{graphicx}
\usepackage{dcolumn}
\usepackage{bm}
\usepackage[linktocpage, colorlinks=true, linkcolor=blue, citecolor=blue]{hyperref}
\usepackage{url}

\usepackage{physics}
\usepackage{xcolor}
\usepackage{mathtools}
\usepackage{amssymb}
\usepackage{color,soul}

\usepackage{enumitem}
\setlist[itemize]{itemsep=1pt, topsep=2pt}

\newcommand{\dbar}{d\hspace*{-0.08em}\bar{}\hspace*{0.1em}}

\begin{document}


\title{Microscopic contributions to the entropy production at all times: \\ From nonequilibrium steady states to global thermalization}

\author{Ayaka Usui}
\email{ayaka.usui@icc.ub.edu}
\email{ayaka.usui@uab.cat}
\affiliation{Dept. F\'{i}sica Qu\`{a}ntica i Astrof\'{i}sica, Institut de Ci\`{e}ncies del Cosmos (ICCUB), Facultat de F\'{i}sica, Universitat de Barcelona, Mart\'{i} i Franqu\'{e}s, 1, E08028 Barcelona, Spain}
\affiliation{Grup d'\`{O}ptica, Departament de F\'{i}sica, Universitat Aut\`{o}noma de Barcelona, Bellaterra (Barcelona) 08193, Spain}

\author{Krzysztof Ptaszy\'{n}ski}
\affiliation{Complex Systems and Statistical Mechanics, Department of Physics and Materials Science, University of Luxembourg, L-1511 Luxembourg, Luxembourg}
\affiliation{Institute of Molecular Physics, Polish Academy of Sciences, Mariana Smoluchowskiego 17, 60-179 Pozna\'{n}, Poland}

\author{Massimiliano Esposito}
\affiliation{Complex Systems and Statistical Mechanics, Department of Physics and Materials Science, University of Luxembourg, L-1511 Luxembourg, Luxembourg}

\author{Philipp Strasberg}
\affiliation{F\'{i}sica Te\`{o}rica: Informaci\'{o} i Fen\`{o}mens Qu\`{a}ntics, Departament de F\'{i}sica, Universitat Aut\`{o}noma de Barcelona, Bellaterra (Barcelona) 08193, Spain}


\begin{abstract}

Based on exact integration of the Schr\"odinger equation, we numerically study microscopic contributions to the entropy production for the single electron transistor, a paradigmatic model describing a single Fermi level tunnel coupled to two baths of free fermions. To this end, we decompose the entropy production into a sum of information theoretic terms and study them across all relevant time scales, including the nonequilibrium steady state regime and the final stage of global thermalization. We find that the entropy production is dominated for most times by microscopic deviations from thermality in the baths and the correlation between (but not inside) the baths. 
Despite these microscopic deviations from thermality, the temperatures and chemical potentials of the baths thermalize as expected, even though our model is integrable.
Importantly, this observation is confirmed for both initially mixed and pure states. We further observe that the bath-bath correlations are quite insensitive to the system-bath coupling strength contrary to intuition. Finally, the system-bath correlation, small in an absolute sense, dominates in a relative sense and displays pure quantum correlations for all studied parameter regimes.

\end{abstract}

\maketitle

\section{Introduction}\label{sec:intro}

Entropy production is a central concept to quantify irreversibility~\cite{KondepudiPrigogineBook2007, SchallerBook2014, fundamental2017Benenti, strasberg2021first, quantum2022Strasberg, GaspardBook2022}. At steady state, for a system coupled to multiple heat baths, it can be expressed in terms of fluxes (e.g., heat flows) and affinities (e.g., temperature differences), which are relatively easy to determine experimentally (e.g., by measuring temperature changes and knowing the heat capacity). However, the precise microscopic mechanisms that are responsible for the production of entropy are masked by this coarse grained perspective. To get a fundamental understanding about the nature, origin and consequences of entropy production, a more detailed perspective is necessary. 

Recently, by building on a microscopic framework for entropy production proposed in Ref.~\cite{entropy2010Esposito} (see also Refs.~\cite{alternative1978Bassett, nonequilibrium1983Lindblad, PeresBook2002, generalization2010Takara, SagawaUedaPRL2010b} for closely related work) such a detailed perspective was provided by two of us~\cite{ptaszynski2019entropy, ptaszynski2020postthermalization, PtaszynskiEspositoPRXQ2023}. By exact numerical integration of a single electron transistor model, i.e., a Fermi level tunnel coupled to two baths of non-interacting fermions, it was observed that the dominant microscopic contribution to the entropy production is given by the correlations between different environmental degrees of freedom (measured in terms of a mutual information precisely defined below). The purpose of the present paper is to consider two important extensions.

\begin{figure}[b]
\centering
\includegraphics[width=.8\linewidth]{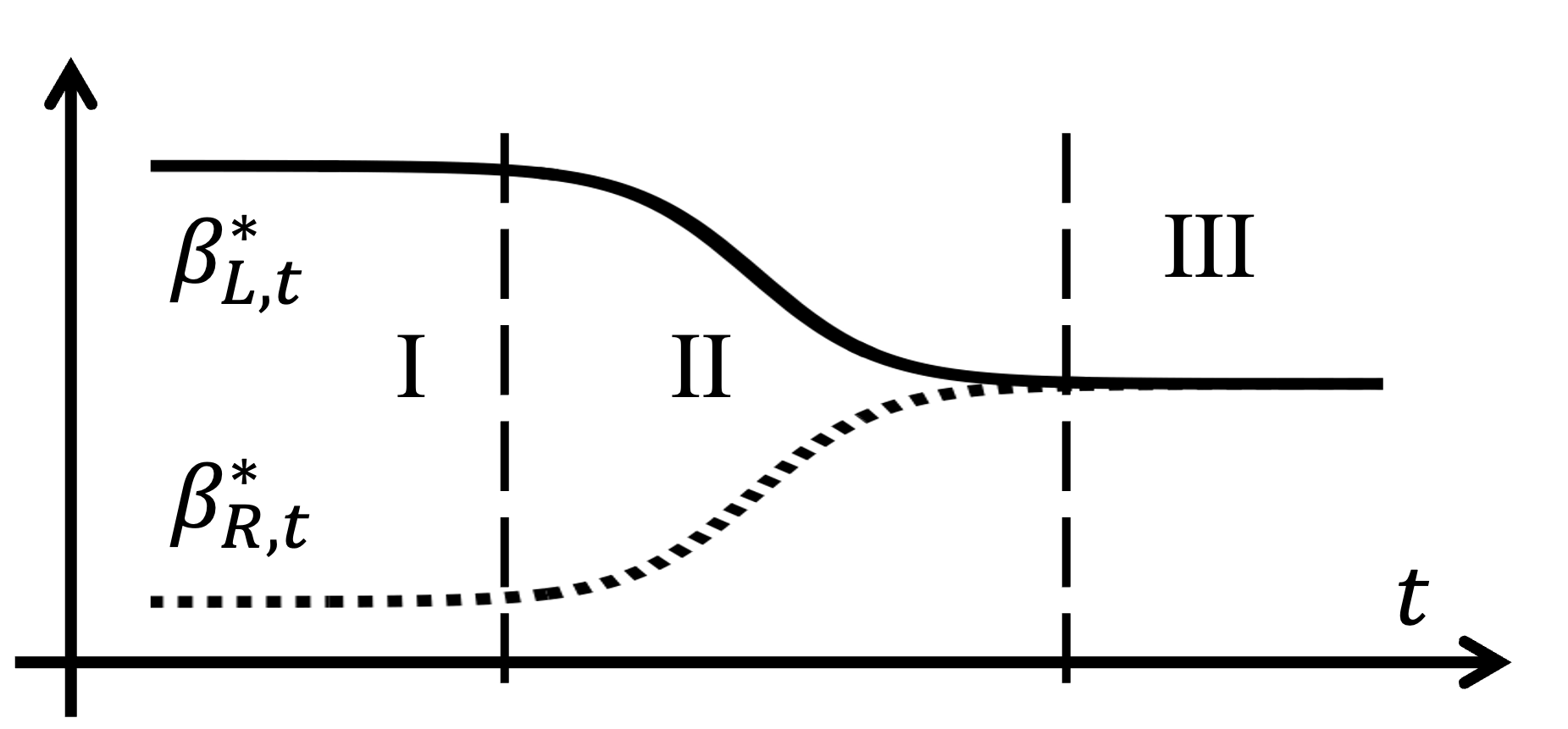}
\caption{
Sketch of how the inverse temperatures $\beta_{L,\tau}^*$, $\beta_{R,\tau}^*$ of the two baths $L$, $R$ change in time (the star $*$ anticipates the fact that we need to define an effective inverse temperature out of equilibrium later on) and how they define the regions I, II and III. While the actual behaviour of the bath temperatures can be more complex because the exchange of electrons can generate a thermoelectric effect, the phenomenology to define the regions I, II and III remains the same. 
}
\label{fig:effbetamu}
\end{figure}

First, by considering a microscopic framework for entropy production valid at all times~\cite{strasberg2021first, strasberg2021clausius, quantum2022Strasberg}, we obtain a temporally complete picture of entropy production starting from the short time dynamics and ending in global thermal equilibrium. To illustrate this, we sketch in Fig.~\ref{fig:effbetamu} the expected macroscopic behaviour in terms of the bath temperatures. During the initial phase, which we call regime I, the baths temperatures hardly change and the system settles into a nonequilibrium steady state. This was the regime studied previously~\cite{ptaszynski2019entropy, ptaszynski2020postthermalization, PtaszynskiEspositoPRXQ2023}. Eventually, due to the flow of heat between the two baths their temperatures start to change slowly, whereas the system adapts on a comparatively fast time scale to new nonequilibrium steady states. We call this regime II. Finally, in the long time limit the heat flow has brought the baths into mutual equilibrium and their temperatures coincide. This signifies global thermalization, here called regime III. Interestingly, owing to the relative size difference between the small system and the large (but finite) baths, we find an exponential time separation between these three regimes. 

From a microscopic perspective we reveal that the entropy production in regime II and III is no longer solely dominated by the correlations inside the environment, in contrast to regime I~\cite{ptaszynski2019entropy, ptaszynski2020postthermalization, PtaszynskiEspositoPRXQ2023}. Instead, microscopic deviations from thermality (precisely quantified by a relative entropy introduced below) become of equal importance. Moreover, we refine the analysis of Refs.~\cite{ptaszynski2019entropy, ptaszynski2020postthermalization, PtaszynskiEspositoPRXQ2023} and divide the correlations inside the environment into correlations contained in each bath separately (called ``intrabath'' correlations) and into the correlations shared between the two baths (called ``interbath'' correlations). We find that the latter are larger than the former. Moreover, while the system-bath correlation cannot significantly contribute to the entropy production in an absolute sense, we find that it is the strongest correlation in a relative sense. In particular, we find pure quantum correlations for all coupling strengths that we study.

The second important extension of our paper is to also consider \emph{pure} initial states on top of mixed initial states with baths described by thermal equilibrium ensembles, which are conventionally considered in open quantum system theory and quantum thermodynamics. Together with the fact that we consider the entire range of time scales, this provides detailed microscopic insights into the emergence of thermodynamic behaviour, from nonequilibrium steady states to global thermalization. In particular, we find that---apart from finite size fluctuations---the ensemble and pure state picture agree well across all time scales. This is surprising and remarkable because our model is integrable, whereas conventional mechanisms of thermalization rely on non-integrability~\cite{DAlessioEtAlAP2016, DeutschRPP2018}. This adds new fuel to the debate whether non-integrability is necessary for the foundations of statistical mechanics, a question which has been addressed for integrable models also in other studies, see, e.g., Refs.~\cite{HuertaRobertsonJSP1969, HuertaRobertsonNearingJMP1971, CramerEisertNJP2010, Lai2015Entanglement,Alba2015Eigenstate,Li2016Quantum,Baldovin2021Statistical, ChakrabortiEtAlJPA2022, HovhannisyanEtAlPRXQ2023}.

It is further worth to note that the here observed phenomenology (as also sketched in Fig.~\ref{fig:effbetamu}) agrees well with transport experiments with cold atoms~\cite{BrantutEtAlScience2012, BrantutEtAlScience2013, HaeuslerEtAlPRX2021}. Assuming the baths to remain in time-dependent grand canonical ensembles, this behaviour was also seen theoretically using linear response theory or master equations~\cite{NietnerSchallerBrandesPRA2014, GallegoMarcosEtAlPRA2014, SchallerNietnerBrandesNJP2014, GrenierGeorgesKollathPRL2014, SekeraBruderBelzigPRA2016, GrenierKollathGeorgesCRP2016}. However, this assumption is in contradiction with microscopic reversible (unitary) dynamics~\cite{VanKampenPhys1954, StrasbergEtAlPRA2023}. Thus, on top of studying microscopic contributions to the entropy production, our paper also microscopically justifies the assumptions used in Refs.~\cite{NietnerSchallerBrandesPRA2014, GallegoMarcosEtAlPRA2014, SchallerNietnerBrandesNJP2014, GrenierGeorgesKollathPRL2014, SekeraBruderBelzigPRA2016, GrenierKollathGeorgesCRP2016} for an experimentally relevant setup.

In the following, we introduce the microscopic decomposition of entropy production in Sec.~\ref{sec:fomulation_sigma} followed by a description of our microscopic model in Sec.~\ref{sec:quantumdots}. Section~\ref{sec:longtime} is the heart of our manuscript and contains our numerical results. Conclusions are drawn in Sec.~\ref{sec:conclusions} and some additional clarifications are relegated to the Appendix.

\section{Decomposition of entropy production}\label{sec:fomulation_sigma}

We consider a quantum system coupled to multiple baths described by the Hamiltonian
\begin{equation}
    \hat{H}_{SB} = \hat{H}_{S} + \hat{H}_{B} + \hat{V}.
\end{equation}
Here, suppressing tensor products with the identity in the notation, $\hat H_S$ denotes the Hamiltonian of the system, $\hat H_B$ the Hamiltonian of the baths and $\hat V$ their interaction with the system. We label the multiple baths by $\nu$ and decompose $\hat{H}_{B}=\sum_{\nu} \hat{H}_{\nu}$. Moreover, in view of our model studied later on, we consider baths that are described by a tensor product of non-interacting subsystems such that we can further decompose $\hat{H}_{\nu}=\sum_{k} \hat{H}_{\nu k}$, where $\hat{H}_{\nu k}$ is the Hamiltonian of the $k$'th mode in bath $\nu$. 

For now, we also consider the following conventional initial states (the discussion of pure initial states is postponed to Sec.~\ref{sec:purestates})
\begin{equation}\label{eq:initial}
    \rho_{SB}(0) = \rho_{S}(0) \otimes \pi_B
    = \rho_{S}(0) \otimes \prod_{\nu}\pi_{\nu}.
\end{equation}
This describes an arbitrary system state $\rho_{S}(0)$ decoupled from baths described by thermal equilibrium ensembles
\begin{align} \label{eq:pi}
    \pi_{\nu} = \pi_\nu(\beta_\nu,\mu_\nu)
    &\equiv
    \frac{\mathrm{e}^{-\beta_{\nu}(\hat{H}_{\nu}-\mu_{\nu}\hat{N}_{\nu})}}{Z_{\nu}}.
\end{align}
Here, $\beta_\nu$, $\mu_\nu$, $\hat N_\nu$ and $Z_\nu$ are the conventional inverse temperature, chemical potential, particle number operator and partition function of the grand canonical ensemble, respectively. In view of the assumed non-interacting nature of the bath, we can write $\pi_{\nu}=\prod_{k}\pi_{\nu,k}$ with obvious definitions for the grand canonical state $\pi_{\nu,k}$ of the $k$'th mode in bath $\nu$. 

Now, for a system exchanging energy and particles with multiple \emph{thermal} baths, i.e., baths that can be well characterized by their energy and particle number (or temperature and chemical potential, respectively), the second law in form of Clausius' inequality stipulates that
\begin{equation}\label{eq:sigma_c}
    \Sigma(t) \equiv \Delta S_{S}(t)
    - \sum_{\nu} \int^t_0 \frac{\dbar Q_{\nu}(\tau)}{T_{\nu}(\tau)}
    \geq 0.
\end{equation}
Here, $\Delta S_S(t) = S_S(t) - S_S(0)$ denotes the change in system entropy, $T_\nu(t)$ is the temperature of bath $\nu$ at time $t$, and $\dbar Q_{\nu}(t) = -[dE_\nu(t) - \mu_\nu(t)dN_\nu(t)]$ is the infinitesimal heat flux \emph{from} bath $\nu$ expressed in terms of its energy change $dE_\nu(t)$ and particle number change $dN_\nu(t)$, the latter multiplied by the chemical potential $\mu_\nu(t)$. The task of this section is then to find microscopic definitions for the (at this point phenomenological) quantities appearing
in Eq.~(\ref{eq:sigma_c}) and to microscopically decompose Eq.~(\ref{eq:sigma_c}) into a sum of non-negative information theoretic terms. 

To this end, we identify the system entropy $S_S(t) = S[\rho_S(t)] = -\mbox{tr}[\rho_S(t)\ln\rho_S(t)]$ with the von Neumann entropy of the system state ($k_B\equiv1$ in the following) and the energy $E_\nu(t) = \mbox{tr}[\hat H_\nu\rho_\nu(t)]$ and particle number $N_\nu(t) = \mbox{tr}[\hat N_\nu\rho_\nu(t)]$ of bath $\nu$ with the expectation value of the respective Hamiltonian and particle number operator. These definitions have been well established for the setup considered here, see, e.g., Refs.~\cite{SchallerBook2014,  strasberg2021first, quantum2022Strasberg, entropy2010Esposito, alternative1978Bassett, nonequilibrium1983Lindblad, PeresBook2002, generalization2010Takara, SagawaUedaPRL2010b, ptaszynski2019entropy, ptaszynski2020postthermalization, PtaszynskiEspositoPRXQ2023, strasberg2021clausius}. Moreover, it has been noticed in Ref.~\cite{strasberg2021first} that the following microscopic definitions of temperature and chemical potential establish the non-negativity of Eq.~(\ref{eq:sigma_c}): 
\begin{align}
    \Tr[\hat{H}_{\nu}\rho_{\nu}(t)] 
    &= \Tr[\hat{H}_{\nu}\pi_{\nu}(\beta^*_{\nu,t},\mu^*_{\nu,t})], \label{eq:betaeff} \\
    \Tr[\hat{N}_{\nu}\rho_{\nu}(t)]
    &= \Tr[\hat{N}_{\nu}\pi_{\nu}(\beta^*_{\nu,t},\mu^*_{\nu,t})]. \label{eq:mueff}
\end{align}
In words, the (inverse) temperature and chemical potential are obtained by equating the actual expectation value of energy and particle number in bath $\nu$, computed with $\rho_\nu(t)$, with the expectations of a grand canonical ensemble $\pi_\nu(\beta^*_{\nu,t},\mu^*_{\nu,t})$ as defined in Eq.~(\ref{eq:pi}). 
Note that this uniquely fixes $\beta^*_{\nu,t}$ and $\mu^*_{\nu,t}$. 
Moreover, we use a star $*$ in the notation to indicate that these are \emph{effective} inverse temperatures and chemical potentials, which are defined for any state $\rho_\nu(t)$. 
We do \emph{not} assume that the true microscopic state of bath $\nu$ is a grand canonical ensemble (indeed, it will be typically far away from it). 

The definition of a nonequilibrium temperature deserves a few comments, given that there is no universal consensus on how to do it~\cite{Casas2003Temperature}. Our definition is not only useful in establishing Clausius inequality, but it has a phenomenological counterpart as explained in Refs.~\cite{Muschik1977Empirical,Muschik1977concept} and shows that it is measurable (at least in principle) even if one has no knowledge about the microscopic Hamiltonian or state of the system. Moreover, since the temperature is functionally dependent only on the average energy, it will eventually agree with the equilibrium temperature if the system can thermalize in the sense of pure state statistical mechanics~\cite{DAlessioEtAlAP2016, DeutschRPP2018}. This was explicitly demonstrated, for instance, in Ref.~\cite{Muller2015Thermalization}. Nevertheless, the temperature used here captures only \emph{one} thermodynamic aspect of the problem: it characterizes the average energy of the system but not other observables. Thus, we believe there is \emph{no} unique physically meaningful definition of temperature for a general nonequilibrium state. For instance, whenever the system displays spatial inhomogeneity, or if the system is in a macroscopic superposition of distinct energies, a single nonequilibrium temperature will not characterize it well. We have focused on discussing temperature only, but similar arguments hold for the chemical potential too.

By identifying $1/T_\nu(t)$ and $\mu_\nu(t)$ in Eq.~\eqref{eq:sigma_c} with $\beta^*_{\nu,t}$ and $\mu^*_{\nu,t}$, respectively, it turns out that we can rewrite Clausius' inequality as
\begin{equation}\label{eq:ClausiusEntropy}
    \Sigma(t) \equiv \Delta S_S(t) + \sum_\nu \Delta\mathcal{S}_\nu(t) \ge 0,
\end{equation}
where $\mathcal{S}_\nu(t) = S[\pi_\nu(\beta^*_{\nu,t},\mu^*_{\nu,t})]$ is the von Neumann entropy of the grand canonical ensemble at the effective inverse temperature and chemical potential. Equation~(\ref{eq:ClausiusEntropy}) nicely illustrates the origin of entropy production arising from an asymmetric treatment of the small system compared to the large bath: whereas the system state is exactly taken into account in the definition of entropy, the only information about the bath that is deemed relevant and accessible from a macroscopic point of view is its energy and particle number (or temperature and chemical potential, respectively). Unfortunately, this still does not reveal which mechanisms at the microscopic level contribute to the entropy production. 

For this purpose, we introduce the non-negative quantum relative entropy $D[\rho\|\sigma] = \mbox{tr}[\rho(\ln\rho-\ln\sigma)]\ge0$ between two states $\rho$ and $\sigma$ and the non-negative quantum mutual information $I_{X-Y} = D[\rho_{XY}|\rho_X\otimes\rho_Y]$ defined for any bipartition $XY$ with state $\rho_{XY}$. Moreover, to avoid notational overload, we write $\pi_{\nu}(t) = \pi_\nu(\beta^*_{\nu,t},\mu^*_{\nu,t})$ and we restrict our attention to the relevant case of two baths labeled $\nu = L$ and $\nu = R$. We then define the following quantities: 
\begin{align}
    & \text{system-bath correlation:} \nonumber \\
    & \qquad I_{S-B}(t) \equiv D[\rho_{SB}(t)|\rho_S(t)\otimes\rho_B(t)], \\
    & \text{interbath correlation:} \nonumber \\
    & \qquad I_{L-R}(t) \equiv D[\rho_{LR}(t)|\rho_L(t)\otimes\rho_R(t)], \\
    & \text{intrabath correlation:} \nonumber \\
    & \qquad I_{\circ\nu}(t) \equiv \sum_k D[\rho_\nu(t)\|\bigotimes_k\rho_{\nu,k}(t)], \\
    & \text{local athermality:} \nonumber \\
    & \qquad D^*_\text{env}(t) \equiv \sum_{\nu,k} D[\rho_{\nu,k}(t)\|\pi_{\nu,k}(t)].
\end{align}
The intrabath correlation $I_{\circ\nu}(t)$ quantifies the correlations \emph{inside} bath $\nu$ shared between the modes $k$, and the local athermality $D^*_\text{env}(t)$ quantifies how far each mode $k$ deviates from the effective grand canonical ensemble introduced above. Note that previously $I_{L-R} + \sum_\nu I_{\circ\nu}(t)$ was considered jointly, not separately~\cite{ptaszynski2019entropy, ptaszynski2020postthermalization, PtaszynskiEspositoPRXQ2023}. Using these definitions and the relation $D[\rho_{XY}||\rho_X\otimes\rho_Y]=S[\rho_X]+S[\rho_Y]-S[\rho_{XY}]$, one can express Eq.~\eqref{eq:sigma_c} as 
\begin{equation}\label{eq:EPdecomposition}
    \Sigma(t) = I_{S-B}(t) + I_{L-R}(t) + \sum_\nu I_{\circ\nu}(t) + D^*_\text{env}(t),
\end{equation}
which, as desired, expresses the entropy production as a sum of non-negative terms, each with a precise information theoretic meaning. 

Finally, it is instructive to compare the formalism above with the framework of Refs.~\cite{entropy2010Esposito, alternative1978Bassett, nonequilibrium1983Lindblad, PeresBook2002, generalization2010Takara, SagawaUedaPRL2010b} where the entropy production is expressed as 
\begin{equation}\label{eq:sigma_d2}
    \sigma(t) \equiv \Delta S_{S}(t) - \sum_{\nu} \frac{Q_{\nu}(t)}{T_\nu(0)} \geq 0.
\end{equation}
In regime I, as studied previously~\cite{ptaszynski2019entropy, ptaszynski2020postthermalization, PtaszynskiEspositoPRXQ2023}, we immediately confirm that $\Sigma(t)\approx\sigma(t)$ as the bath temperatures hardly change: $T_\nu(t) \approx T_\nu(0)$. In regime II and III, however, $\Sigma(t)$ and $\sigma(t)$ can deviate significantly, and their difference is universally given by (even if the initial bath states are not in thermal equilibrium)~\cite{strasberg2021clausius}
\begin{equation} \label{eq:D_env_eq}
    \sigma(t) - \Sigma(t) 
    = \sum_{\nu,k} D[\pi_{\nu,k}(t)||\pi_{\nu,k}] \equiv D_\text{env}^\text{eq}(t) \geq 0.
\end{equation}
We call $D_\text{env}^\text{eq}(t)$ the finite bath effect. It quantifies how much the \emph{thermal} properties of the bath change in time.

\section{Model System}\label{sec:quantumdots}

\begin{figure}[tb]
\centering
\includegraphics[width=.95\linewidth]{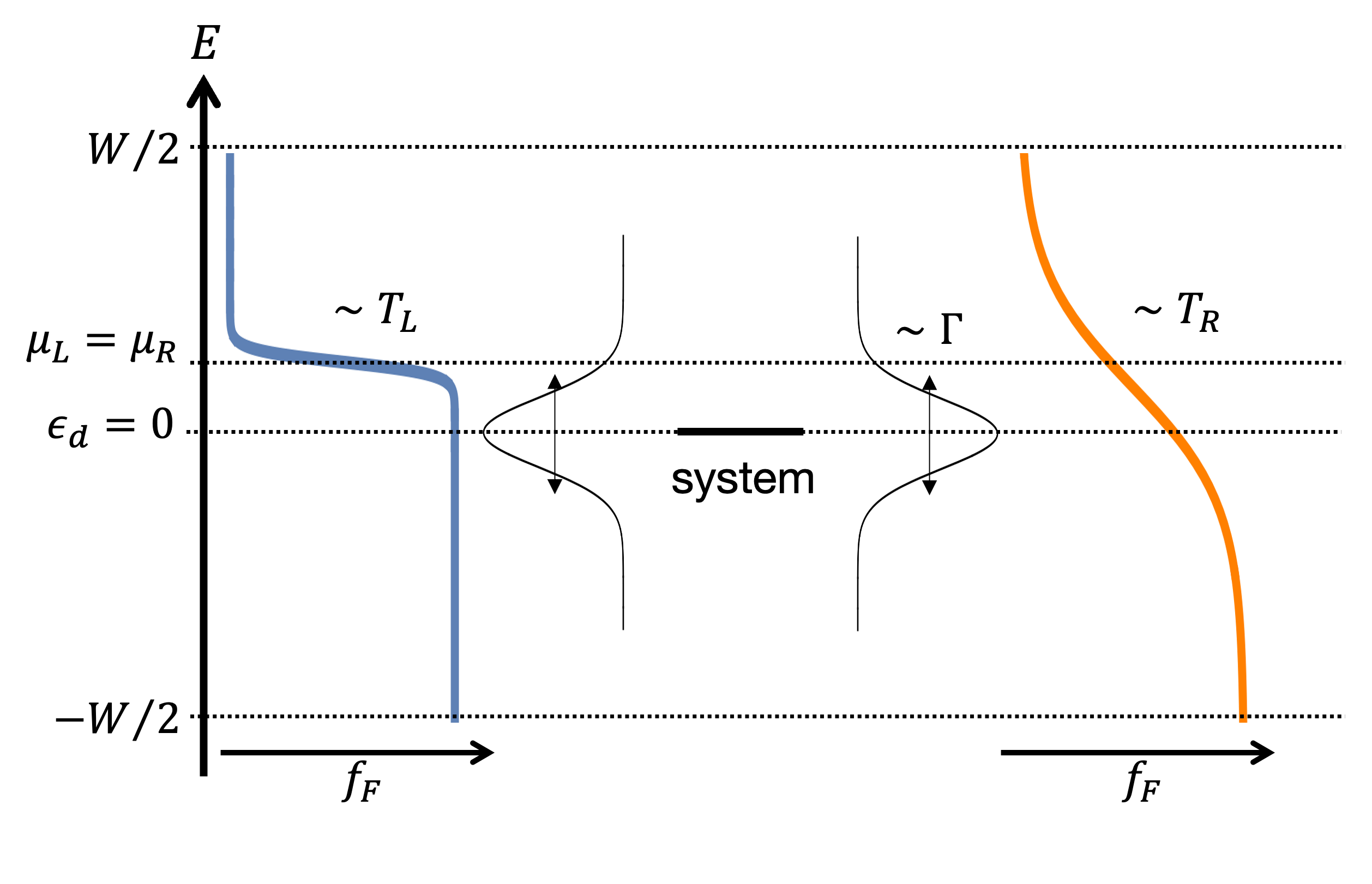}
\caption{
Sketch of the setup. The system at the centre consists of one level and each bath consists of $K$ equally spaced modes (not sketched) in the energy window $W$. We sketch the level broadening proportional to the coupling strength $\Gamma_\nu$. The particle distributions on the left and right baths are given by the Fermi distribution with chemical potential $\mu_\nu$ and temperature $T_\nu$ in bath $\nu$. Note that the true state of the bath for times $t>0$ is not a grand canonical ensemble.
}
\label{fig:systemLR}
\end{figure}

We consider the single electron transistor or single resonant level model, an effective model widely used to study quantum transport processes in condensed matter~\cite{HaugJauho1996, NazarovBlanterBook2009, Paulsson2012Resistance, SchallerBook2014, quantum2022Strasberg}. It has been also used to describe a variety of experiments, including a recent realization as a thermoelectric generator~\cite{JosefssonEtAlNatNano2018}. It is described by the following Hamiltonian with $c^{(\dagger)}$ denoting conventional fermionic annihilation (creation) operators: 
\begin{align}
    \hat{H}_S &= \epsilon_d \hat{c}^{\dagger}_d\hat{c}_d, \\
    \hat H_B &= \sum_{\nu\in\{L,R\}}\sum_{k=1}^K \epsilon_{\nu k} \hat{c}^{\dagger}_{\nu k}\hat{c}_{\nu k}, \\
    \hat V &= \sum_{\nu\in\{L,R\}}\sum_{k=1}^K\left(t_{\nu k} \hat{c}^{\dagger}_{d}\hat{c}_{\nu k}    +\text{H.c}\right).
\end{align}
We take the $K$ energy levels of the bath $\nu$ to be equally spaced, $\epsilon_{\nu k}=(k-1)\Delta\epsilon-W/2$, with $W$ the energy window and $\Delta \epsilon=W/(K-1)$ the level spacing. Moreover, we consider the common wide-band limit, which assumes the tunnel matrix elements $t_{\nu k}=t_{\nu}^0\equiv\sqrt{\Gamma_{\nu}\Delta \epsilon/2\pi}$ to be independent of $k$ with $\Gamma_{\nu}$ the coupling strength to the bath $\nu$. Such a structureless bath corresponds most closely to Markovian behaviour, but we remark that structured baths can be included into the description by using Markovian embedding strategies~\cite{WoodsEtAlJMP2014, StrasbergEtAlPRB2018, SchallerNazirBook2018}. The result will be a slightly increased system consisting of several fermionic levels coupled to a residual structureless bath. Thus, as long as the baths are large compared to the system, it is reasonable to expect the same qualitative behaviour as reported below. Furthermore, to make transport efficient, we need to choose parameters such that the levels in the bath couple well to the system. This can be formalized by demanding that the level broadening in the system~\cite{Paulsson2012Resistance} overlaps with sufficiently many levels in the bath, i.e., we need to choose $\Delta\epsilon\ll\Gamma_{\nu}$. 
The resulting model is sketched in Fig.~\ref{fig:systemLR}. 

To compute physical quantities, we need to solve the Schr\"odinger equation of the model. Since the model is quadratic and the initial state Gaussian, the full many-body density matrix can be encoded into the correlation matrix $\mathcal{C}(t)$~\cite{Ingo2003Calculation,Eisler2012entanglement,sharma2015landauer}, whose elements are given by
\begin{equation}
    \mathcal{C}_{ij}(t) = \Tr[\hat{c}^{\dagger}_{i}\hat{c}_{j} \rho_{SB}(t)].
\end{equation}
Note that this is only a $(2K+1)\times(2K+1)$ matrix in contrast to the full $2^{2K+1}\times2^{2K+1}$ many-body density matrix. The time evolution of the correlation matrix is calculated as
\begin{equation} \label{eq:matCt}
    \mathcal{C}(t) = 
    \mathrm{e}^{i\mathcal{H}t}\mathcal{C}(0)\mathrm{e}^{-i\mathcal{H}t},
\end{equation}
where the elements of the Hamiltonian matrix $\mathcal{H}$ are $\mathcal{H}_{ii}=\epsilon_{ii}$ and $\mathcal{H}_{\nu k,d}=\mathcal{H}_{d,\nu k}=t_{\nu k}$.
Since the initial baths states are thermal states, the initial correlation matrix $\mathcal{C}(0)$ is diagonal and given by
\begin{align}
    \mathcal{C}(0)
    &=
    \mathrm{diag}
    [n_d, n_{L1}, \ldots, n_{LK}, n_{R1}, \ldots, n_{RK}]
    ,
\end{align}
where $n_d$ is the initial occupancy at the system and $n_{\nu k}=f(\beta_{\nu} (\epsilon_{\nu k}-\mu_{\nu}))$  is the Fermi distribution of mode $k$ in bath $\nu$. 

\begin{figure}[t]
\centering
\includegraphics[width=.95\linewidth]{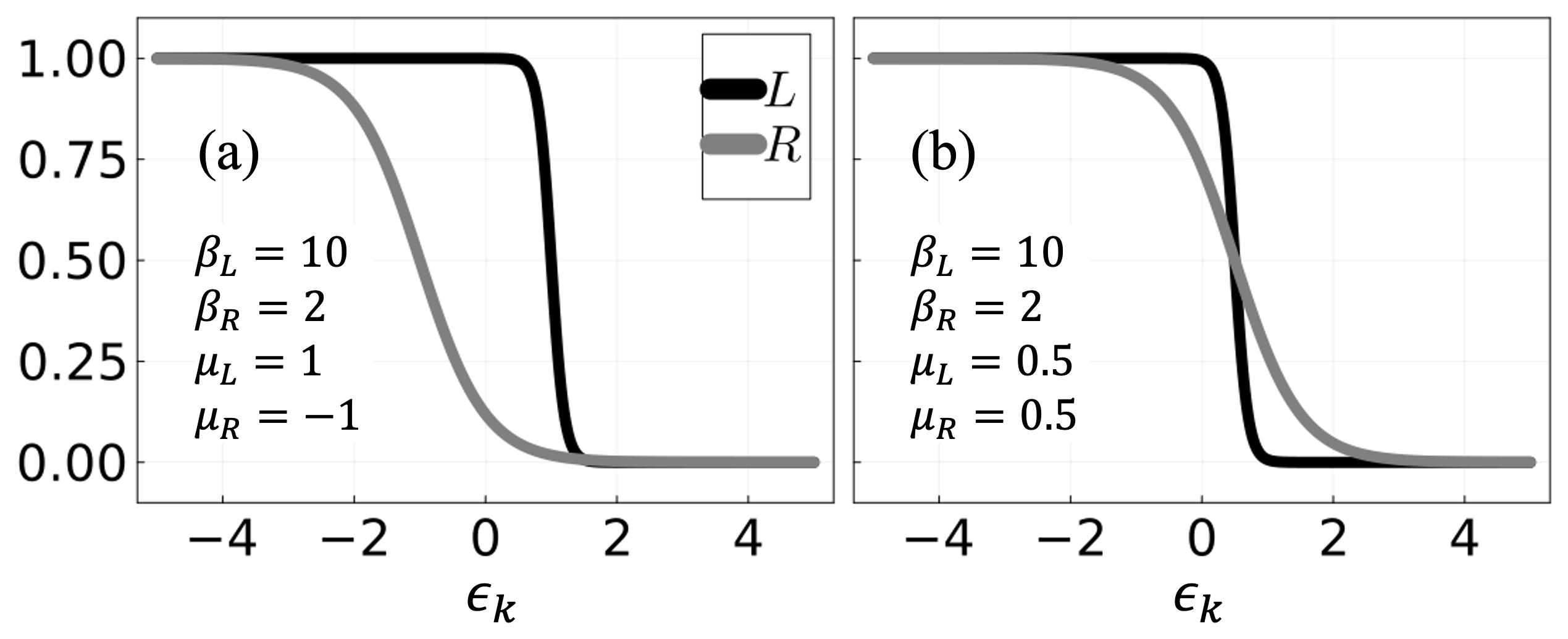}
\caption{
Initial Fermi distributions used for numerical calculation in case (a) and (b).
}
\label{fig:initial}
\end{figure}

The von Neumann entropy of the subspace $\mathcal{G}\in\{S,L,R\}$ is computed as~\cite{sharma2015landauer}
\begin{align}
    S_{\mathcal{G}}
    &=
    -\sum_{\alpha}\left[
    \lambda_{\alpha} \log \lambda_{\alpha}
    +
    \left(1-\lambda_{\alpha}\right) \log \left(1-\lambda_{\alpha}\right) 
    \right]
    ,
\end{align}
where $\lambda_{\alpha}$ is the eigenvalues of the reduced correlation matrix in the subspace $\mathcal{G}$. This enables us to calculate the correlations $I_{S\text{-}B}(t)$, $I_{L\text{-}R}(t)$, $I_{\circ\nu}(t)$. The heat flux can be expressed in terms of the correlation matrix as
\begin{align}
    \dbar Q_{\nu}(t)
    &= \sum_{k}\left[
    \mathcal{C}_{\nu k,\nu k}(t) - \mathcal{C}_{\nu k,\nu k}(t+dt)
    \left(
    \epsilon_{\nu k} - \mu_{\nu}
    \right)
    \right]
    .
\end{align}
The effective temperatures and chemical potential can be extracted by directly solving Eqs.~(\ref{eq:betaeff}) and~(\ref{eq:mueff}) or, alternatively, by computing the heat capacity matrix $C$ of bath $\nu$ in the grand canonical ensemble and by inverting $(dE,dN)^T = C(dT,d\mu)^T$. Once the effective grand canonical ensemble is obtained, we can compute the local athermality $D_{\text{env}}^*(t)$ and the finite bath effect $D_\text{env}^\text{eq}(t)$.

\section{Numerical results}\label{sec:longtime}

\begin{figure*}[bt]
\centering
\includegraphics[width=0.95\linewidth]{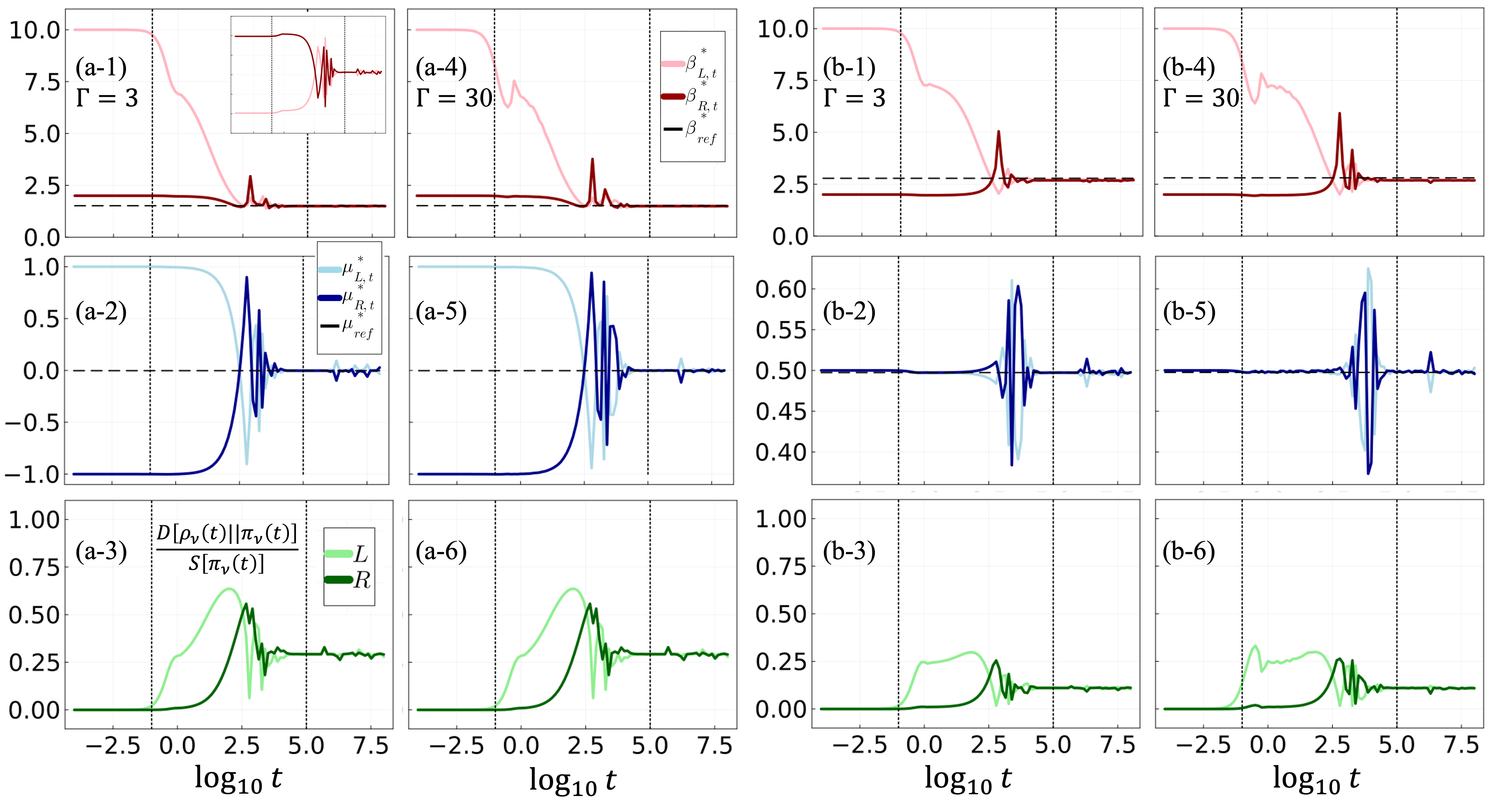}
\caption{From top to bottom: Time evolution of the inverse temperature, the chemical potential and the deviation of the true microscopic state from the corresponding grand canonical ensemble (precisely defined in the text). Note the logarithmic time scale. The first two columns correspond to case (a) with $\Gamma = 3$ (a-1,a-2,a-3) or $\Gamma = 30$ (a-4,a-5,a-6). The last two columns correspond to case (b) with $\Gamma = 3$ (b-1,b-2,b-3) or $\Gamma = 30$ (b-4,b-5,b-6). Note that the quantitative behaviour is quite insensitive to the different chosen $\Gamma$. Moreover, in the first two rows the dashed line corresponds to the temperature or chemical potential of a \emph{global} grand canonical state as explained in the text. The inset in subfigure (a-1) sketches the change in energy (instead of temperature) for reasons explained in the text. Finally, we use vertical lines to delimit regime I, II and III. All other numerical parameters are detailed at the beginning of Sec.~\ref{sec:longtime}.
}
\label{fig:efftemchemi}
\end{figure*}

We now study the full time behaviour of entropy production and its microscopic components numerically. 
The code we have used is available online~\footnote{\href{https://zenodo.org/doi/10.5281/zenodo.10539085}{https://zenodo.org/doi/10.5281/zenodo.10539085}}.
We always set $\epsilon_{d}=0$ and assume an initially empty system state, $n_d = 0$. Also, we fix the initial inverse temperatures to $\beta_L=10$, $\beta_R=2$ but consider the following two cases of initial chemical potentials,
\begin{itemize}
    \item[(a)] $\mu_{L}=-\mu_{R}=1.0$, 
    \item[(b)] $\mu_{L}=\mu_{R}=0.5$.
\end{itemize} 
In both cases the baths have different energies, but only in case (a) the baths have different particle numbers. 
Figure~\ref{fig:initial} (a) and (b) display the initial Fermi distributions for the two different cases and show that in (a) there is a wide range of energy levels that are fully occupied in bath $L$ but not in bath $R$ while in (b) there is room for particles to hop in or out only near Fermi level.
To satisfy the condition $\Delta\epsilon\ll\Gamma_{\nu}$, we choose two identical bath Hamiltonians with $K=10^3$ levels, energy window $W=10$ (resulting in $\Delta\epsilon=0.01$) and a (symmetric) coupling strength $\Gamma_L=\Gamma_R=\Gamma\geq 3$. If the coupling strength falls below $\Gamma = 3$, the levels do not interact well, and energy and particle exchange is hindered. This becomes particularly clear when considering \emph{asymmetric} baths, which we briefly investigate in Appendix~\ref{app:asymmetry}. 

We remark that coupling strengths $\Gamma\ge3$ are rather large from the perspective of the open quantum system: a weak coupling master equation for the system would not describe well the transport process. However, seen from a global perspective, we are at weak coupling because the bath energies dominate typical energy fluctuations in the interaction Hamiltonian. Thus, in some sense, while the open system is not weakly coupled, the baths are. This latter (and less stringent) notion of weak coupling is sufficient for our purposes.

\subsection{Effective nonequilibrium temperature and chemical potential} \label{sec:eff_temp_chemipo}

Before turning to the study of entropy production, it is instructive to consider the evolution of the effective (inverse) temperatures and chemical potentials (from now on we drop the word ``effective'' for brevity). To this end, we not only distinguish between case (a) and (b), but also consider two different coupling strengths $\Gamma=3$ and $\Gamma = 30$. The results are summarized in Fig.~\ref{fig:efftemchemi}. Based on the numerical evidence we draw the following conclusions. 

First, as anticipated at the beginning and sketched in Fig.~\ref{fig:effbetamu}, the three different regimes I, II and III become clearly visible. To repeat, they are characterized by the intensive thermodynamic variables (temperature and chemical potential) staying initially constant (I), changing (II) and settling into a new final value (III). Moreover, note the exponential separation of time scales between the regimes. 

Second, we observe that in regime III the intensive thermodynamic variables settle into a \emph{common} value, characterizing global thermal equilibrium (thermalization). This, of course, is the expected behaviour from macroscopic thermodynamics, but note that it emerges here in a unitarily evolving system for effective temperatures and chemical potentials defined for states that are out of equilibrium from a microscopic perspective.

The claim that we observe true global thermalization is further supported by the dashed grey lines in Fig.~\ref{fig:efftemchemi}. These lines are obtained by extracting the inverse temperature $\beta^*_\text{ref}$ and chemical potential $\mu^*_\text{ref}$ from a \emph{global} grand canonical ensemble $\pi_{SB} \sim \exp[-\beta^*_\text{ref}(\hat H_{SB}-\mu^*_\text{ref} \hat N_{SB})]$ (which includes the system and the system-bath interaction) by considering the conserved expectation values of the global energy $\langle\hat H_{SB}\rangle$ and particle number $\langle\hat N_{SB}\rangle$. Details on how to compute the global grand canonical ensemble are given in Appendix~\ref{app:globalgibbs}. This demonstrates thermalization from a macroscopic point of view in an integrable model.

We remark that the observed global thermalization is not an artefact from choosing identical (symmetrical) baths with the same number of modes $K$ and energies $\epsilon_{\nu k}$. In fact, we confirmed that our conclusions are robust for asymmetric baths with different numbers of modes, as indicated in Appendix~\ref{app:asymmetry}. 

We further observe a rather counter-intuitive behaviour of the inverse temperatures in case (a). Instead of settling into a common value $\beta^*_\text{ref}$ that lies \emph{between} the initial values $\beta_{L,0}^*$ and $\beta_{R,0}^*$, \emph{both} baths heat up and settle to a lower inverse temperature. This is a consequence of the bath energy depending on both the temperature and chemical potential in a rather complicated algebraic way. The inset in Fig.~\ref{fig:efftemchemi} (a-1) sketches the evolution of the bath energies and shows the expected behaviour: the final energy lies between the initial energies.

Third, we also observe a clear transient thermoelectric effect in case (b), where the initial temperature imbalance creates an imbalance in chemical potential before reaching some final global equilibrium state. This is, of course, an expected effect of the single electron transistor~\cite{HaugJauho1996, NazarovBlanterBook2009, Paulsson2012Resistance, SchallerBook2014, fundamental2017Benenti, quantum2022Strasberg, JosefssonEtAlNatNano2018} and such a transient thermoelectric effect was precisely observed in transport experiments with cold atoms~\cite{BrantutEtAlScience2012, BrantutEtAlScience2013, HaeuslerEtAlPRX2021}. This demonstrates that our approach provides at least a minimal qualitative model to describe such experiments (cf.~Refs.~\cite{NietnerSchallerBrandesPRA2014, GallegoMarcosEtAlPRA2014, SchallerNietnerBrandesNJP2014, GrenierGeorgesKollathPRL2014, SekeraBruderBelzigPRA2016, GrenierKollathGeorgesCRP2016}).

\begin{figure*}[tb]
\centering
\includegraphics[width=.95\linewidth]{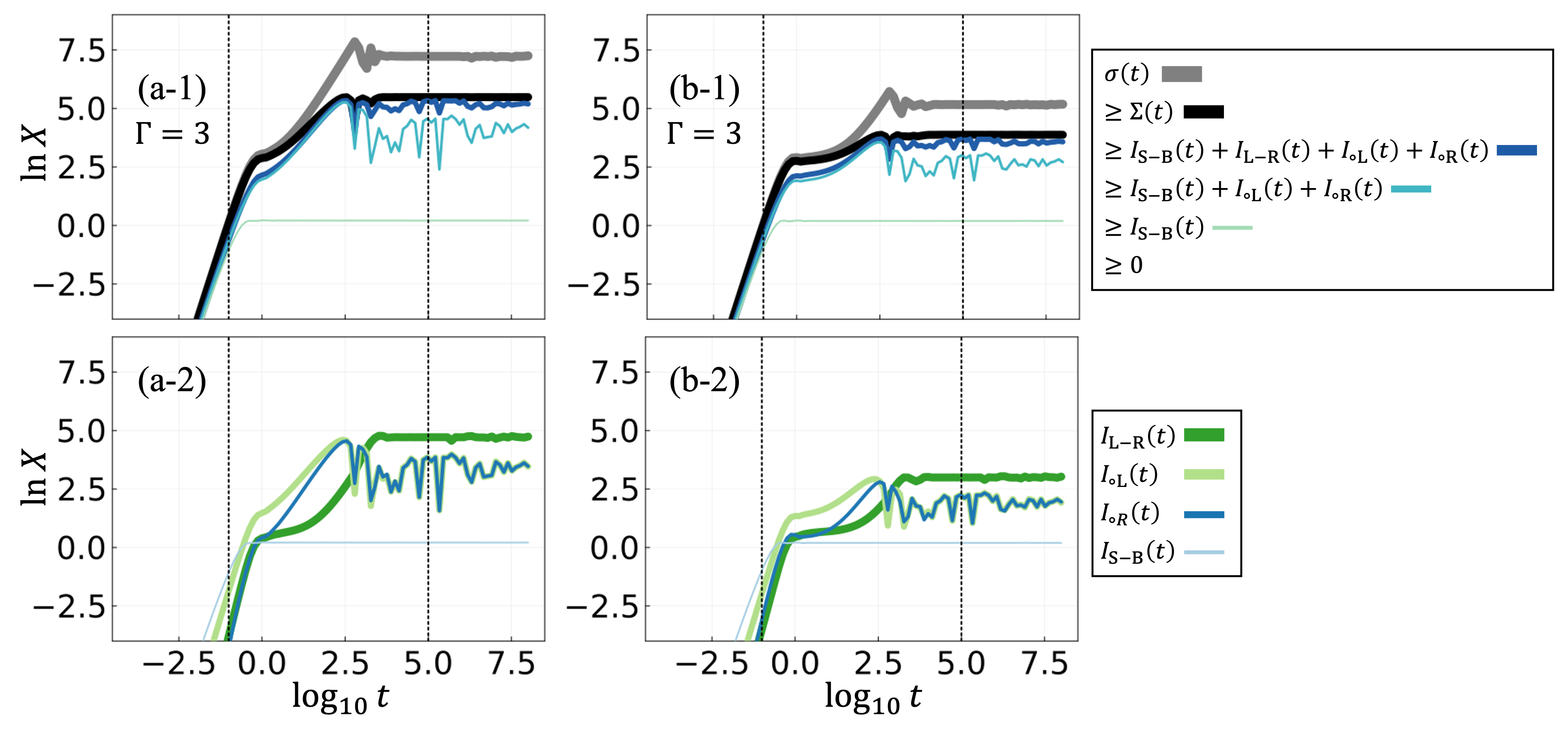}
\caption{ 
Top row: Time evolution of the entropy production $\sigma(t)$ and $\Sigma(t)$ and their components ordered in a hierarchical way as detailed in the legend to the right. Bottom row: Time evolution of the different mutual information terms contributing to the entropy production. Notice that we plot $I_{\circ L}$ and $I_{\circ R}$ separately in contrast to the first row. The first column refers to case (a) with $\Gamma=3$ and the second column refers to case (b) with $\Gamma=3$. The remaining parameters are chosen as before. Notice that \emph{both} axes have a logarithmic scale.
}
\label{fig:sigma}
\end{figure*}

Fourth and finally, the last row of Fig.~\ref{fig:efftemchemi} demonstrates how far the actual microscopic state of the baths deviates from the corresponding grand canonical state obtained by matching the temperature and chemical potential via Eqs.~(\ref{eq:betaeff}) and~(\ref{eq:mueff}). To find a good distance measure, we notice that $D[\rho_{\nu}(t)||\pi_{\nu}(t)] = -S[\rho_{\nu}(t)] + \mathcal{S}_\nu(t)$ with the von Neumann entropy $\mathcal{S}_\nu(t)$ of the grand canonical ensemble as introduced below Eq.~(\ref{eq:ClausiusEntropy}). Thus, we have $D[\rho_{\nu}(t)||\pi_{\nu}(t)]/\mathcal{S}_\nu(t) \in [0,1]$, which makes it a good distance measure (note that the upper bound cannot be reached for mixed bath states). From Fig.~\ref{fig:efftemchemi} we then infer that the bath remains close to the grand canonical state in regime I as expected, whereas it can markedly differ from it in regimes II and III, clearly indicating that our results are not restricted to the linear response regime of small biases in temperature and chemical potential. We further observe a non-monotonic behaviour, i.e., the microscopic bath states are \emph{not} continuously pushed away from the corresponding equilibrium state. 
Moreover, it is remarkable that also this abstract information theoretic distance measure ``equilibrates'' for both baths to the \emph{same} value in regime III, an observation that is certainly not required by the laws of thermodynamics. At present, however, it remains unclear how generic this effect is. Finally, notice that the quantity we plot is closely related to the terms contributing to the entropy production via the identity $\sum_{\nu=L,R}D[\rho_{\nu}(t)||\pi_{\nu}(t)]=\sum_{\nu=L.R}I_{\circ \nu}(t)+D_{\text{env}}^*(t)$. 

\subsection{Entropy production and its components}

We now turn to the study of entropy production and its components as identified in Eq.~(\ref{eq:EPdecomposition}), i.e., we still restrict the discussion to the initial ensemble given in Eq.~(\ref{eq:initial}). Moreover, for comparison we also study the quantity defined in Eq.~(\ref{eq:sigma_d2}), which coincides with the entropy production for baths with constant temperature and chemical potential. 

First insights can be gained from Fig.~\ref{fig:sigma} where we plot all quantities as a function of time for case (a) and (b) and $\Gamma=3$. To understand the full temporal evolution, we use a logarithmic scale not only for time but also for entropy production and its components. Nevertheless, it is not possible to resolve in detail all terms on all time scales owing to the smallness and quick growth in regime I. However, since regime I is well covered in Refs.~\cite{ptaszynski2019entropy, ptaszynski2020postthermalization, PtaszynskiEspositoPRXQ2023}, we decided to focus on regimes II and III. In regime II we can see that a clear hierarchy between different terms emerges (recall the logarithmic scale), which we discuss in more detail below. Instead, in regime III all quantities settle to a stationary value, although it is interesting to observe that the intrabath correlations $I_{\circ\nu}$ display strong relative fluctuations around their mean value. In contrast, the system-bath correlation $I_{S-B}$ settles to an almost perfectly constant value already at the very early stages in regime II. Moreover, while the interbath correlation $I_{L-R}(t)$ as well as $\Sigma(t)$ and $\sigma(t)$ also fluctuate, their fluctuations are relatively small compared to their mean. This behaviour is further exemplified in the second row of Fig.~\ref{fig:sigma}, which also demonstrates that, despite the fluctuations, $I_{\circ L}(t)\approx I_{\circ R}(t)$ in regime III. Again, similar to the third row in Fig.~\ref{fig:efftemchemi}, the observation that both baths show the same information theoretic behaviour goes beyond what one would expect from textbook thermodynamic behaviour (which only predicts that the intensive variables reach the same value).

\begin{figure}[tb]
\centering
\includegraphics[width=.95\linewidth]{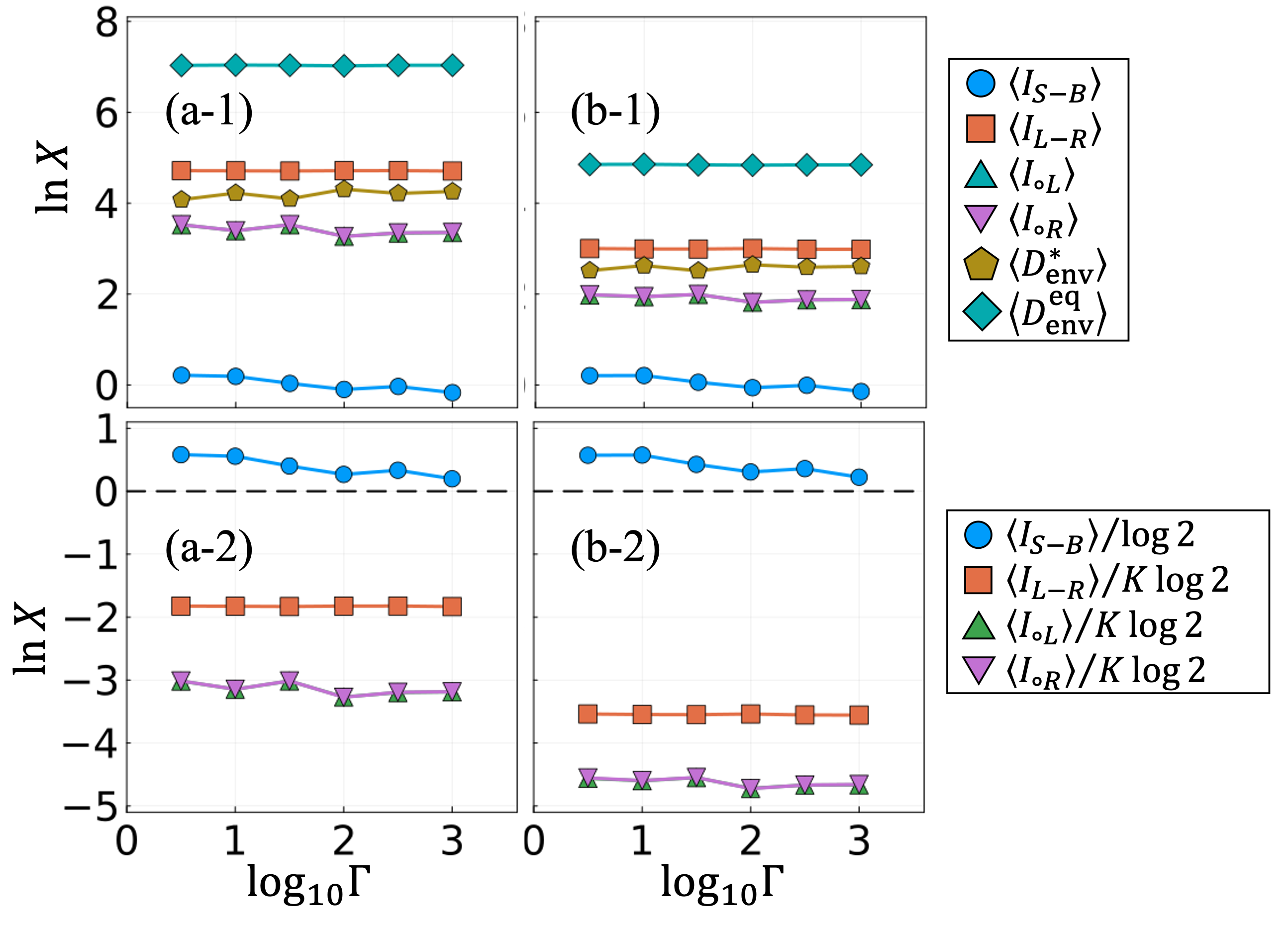}
\caption{Top row: Plot of the different components contributing to the final value of the entropy production in regime III as a function of the coupling strength. To filter out fluctuations, we average all quantities from $\Gamma t=10^5$ to $\Gamma t=10^8$. Bottom row: Plot of the correlations normalized by their maximum classical value. Note that the system-bath correlations always exceeds the classical bound (grey dashed line). First column: Case (a). Second column: Case (b). Note the double logarithmic scale on all plots.
}
\label{fig:correlations}
\end{figure}

To study the influence of the different components on the final behaviour of the entropy production in detail, we plot each component in regime III in Fig.~\ref{fig:correlations} by varying the coupling strength $\Gamma$ over three orders of magnitude. The first important and remarkable observation is that all quantities are fairly constant as a function of coupling strength, allowing us to claim a certain universality for our results, at least with respect to the single electron transistor. Next, we see that the by far largest term is the finite bath effect $D_\text{env}^\text{eq}(t)$, showing that $\sigma(t)$ overestimates the true entropy production $\Sigma(t)$ by many orders of magnitude, which has interesting consequences for the study of heat engines~\cite{strasberg2021clausius}. The most dominant contribution to $\Sigma(t)$ is roughly equally shared by the interbath correlation $I_{L-R}(t)$ and the local athermality $D^*_\text{env}(t)$, whereas the intrabath correlations $I_{\circ\nu}(t)$ lack a bit behind (it should be noted, however, that the sum of the local athermality and the intrabath correlations add up to global athermality). Thus, in contrast to regime I, it is no longer the case that the entropy production is dominated by the inter- and intrabath correlations~\cite{ptaszynski2019entropy, ptaszynski2020postthermalization, PtaszynskiEspositoPRXQ2023}. Instead, (local) athermality gives rise to a non-negligible contribution too, which perhaps is not surprising in regime III where the baths are macroscopically different from regime I. 

While Fig.~\ref{fig:sigma} has already shown that the system-bath correlations $I_{S-B}(t)$ give rise to a negligible contribution in an absolute sense, the bottom row of Fig.~\ref{fig:correlations} demonstrates that they represent the strongest contribution from a relative point of view. To see this, we use that classical mutual information between two systems $X$ and $Y$ is upper bounded by $I_{X-Y} \le \min\{d_X,d_Y\} \equiv I_\text{cl}^\text{max}$, where $d_X$ ($d_Y$) denotes the number of states in $X$ ($Y$). In the bottom row of Fig.~\ref{fig:correlations} we then plot the various normalized correlations $I/I_\text{cl}^\text{max}$, where $I$ denotes the quantum mutual information. Interestingly, quantum correlations can be potentially twice as strong as classical correlations, i.e., $I\le 2I_\text{cl}^\text{max}$, and we observe that for all coupling strength $I_{S-B}(t)$ exceeds the classical bound. This signifies the presence of quantum correlations, which are caused by our choice of relatively low temperatures.

We remark that the relatively strong system-bath $I_{S-B}(t)$ and interbath correlations $I_{L-R}(t)$ indicate that the full density matrix cannot be approximated by a decorrelated state. Nevertheless, since the baths thermalize as expected, they behave from a macroscopic point of view \emph{as if} they were described by decorrelated grand canonical ensembles.


\subsection{Entropy production for pure states}\label{sec:purestates}

\begin{figure}[b]
\centering
\includegraphics[width=.7\linewidth]{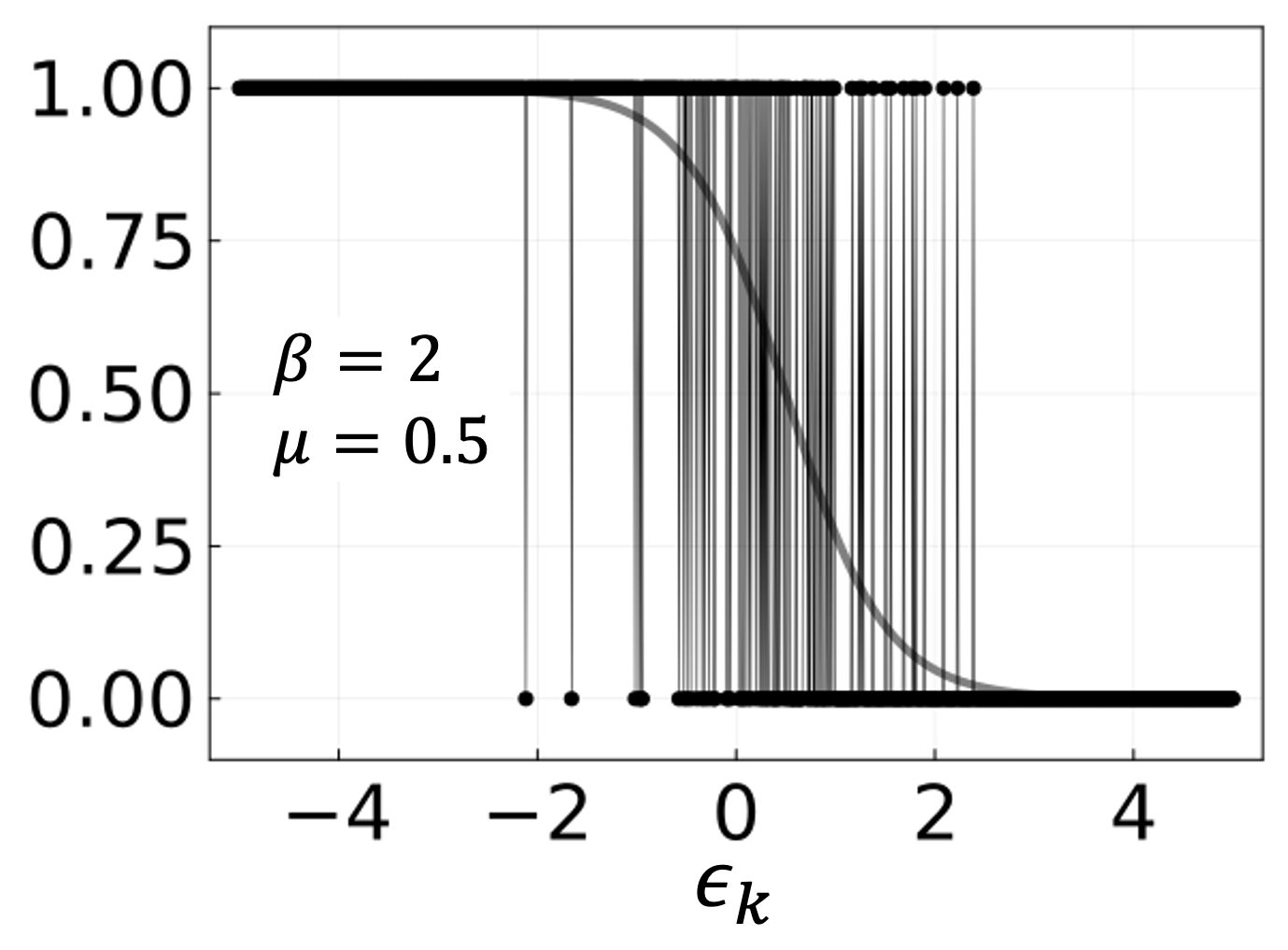}
\caption{
Example of an initial pure state used for numerical calculation. The black dots represent the population of the pure state, and the grey line represents the corresponding Fermi distribution. 
}
\label{fig:initial_pure}
\end{figure}

In the final section of our paper we turn to a key question in statistical mechanics, namely how sensitive our conclusions depend on the precise choice of the initial state. Indeed, statistical physics teaches us that the vast majority of microscopic states that look initially the same from a macroscopic point of view should also give rise to the same macroscopic behaviour~\cite{GemmerMichelMahlerBook2004}. This then justifies to replace the actual state by a (fictitious) ensemble. Conventionally, the emergence of Gibbsian ensemble statistical mechanics is explained using non-integrability~\cite{DAlessioEtAlAP2016, DeutschRPP2018}, but several indications exist that non-integrability might not be essential~\cite{HuertaRobertsonJSP1969, HuertaRobertsonNearingJMP1971, CramerEisertNJP2010, Baldovin2021Statistical, ChakrabortiEtAlJPA2022, HovhannisyanEtAlPRXQ2023,Li2018Quantum, Lai2015Entanglement,Alba2015Eigenstate,Li2016Quantum}. The evidence presented below confirms the latter point. 

To approach the problem, we first need valid candidates for pure states that accurately reflect the macroscopic knowledge we have about the initial energy and particle number of the baths. To use our formalism (see Sec.~\ref{sec:quantumdots}), we need to choose an initial state that is Gaussian. This can be constructed by randomly sampling the Fermi distribution as indicated in Fig.~\ref{fig:initial_pure}. This is done by drawing independent uniform random numbers $r\in[0,1]$ for each mode $k$ of each bath $\nu$ and by (not) populating the level if $r$ is below (above) the Fermi function. Thus, our initial pure state is an energy eigenstate of the local Hamiltonian $\hat{H}_S+\hat{H}_B$ with a definite particle number. Moreover, by construction we obtain our initial state in Eq.~(\ref{eq:initial}) on average. We then run the dynamics as before and compute the same quantities, again for case (a) and (b) and $\Gamma=3$. 

\begin{figure}[tb]
\centering
\includegraphics[width=.95\linewidth]{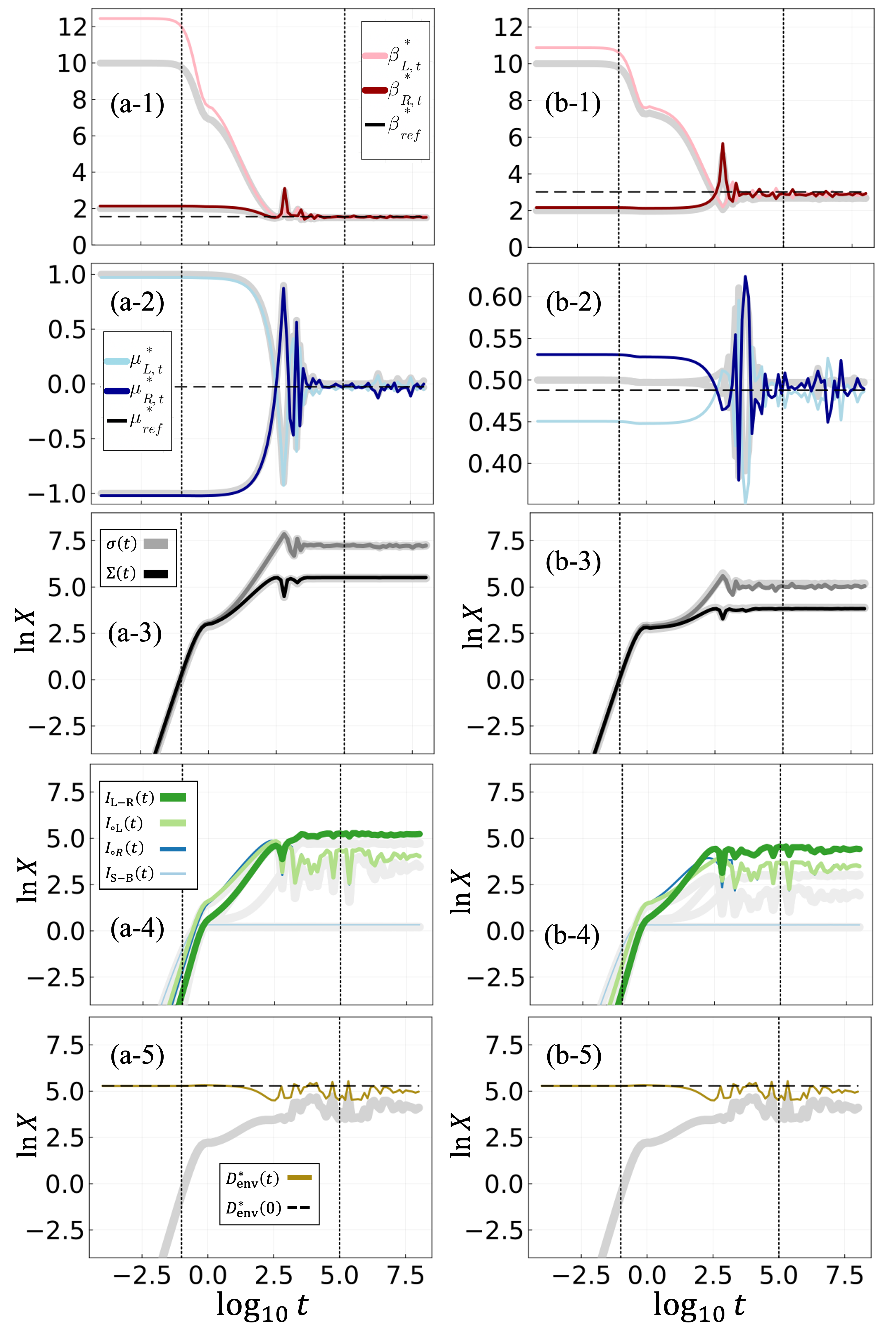}
\caption{
First column from top to bottom: Time evolution of the temperature, chemical potential, entropy production, the correlations and the local athermality as a function of time for case (a) and $\Gamma=3$. Second column from top to bottom: The same for case (b). The thick grey lines correspond to the previous data starting from the initial ensemble in Eq.~(\ref{eq:initial}). The coloured, dark, thin lines correspond to a single realization of a randomly drawn pure state as explained in the text.
}
\label{fig:purestateb}
\end{figure}

The results are shown in Fig.~\ref{fig:purestateb}, where we compare the time evolution starting from a randomly drawn pure state (coloured, dark, thin lines) with the previous ensemble averaged results (thick, grey lines). We find it very remarkable to observe a close quantitative agreement of all thermodynamic quantities (temperature, chemical potential, entropy production) across all time scales. Some discrepancies exist, but they could well be explained by finite size corrections, which, among other things, cause an initial slight mismatch in temperature and chemical potential (or energy and particle number, respectively) in each realization. Systematic (and expected) deviations were only found for the microscopic components of the entropy production, see Fig.~\ref{fig:purestateb} (a-4,a-5) and (b-4,b-5), to which we turn below. This is consistent to Ref.~\cite{Ptaszynski2023Ensemble}, where similar results were observed for initial microcanonical (instead of grand canonical) states.

Importantly, we observed these results to be \emph{typical}, i.e., different random realizations of the initial state give rise to approximately the same behaviour (not displayed for brevity). This indicates that the results in Fig.~\ref{fig:purestateb}, and the conclusions we have drawn in general, are robust under a change of the initial state (at least for the class of initial states considered here).

Finally, let us study the microscopic contributions to the entropy production in more detail. We start from Clausius' inequality, Eq.~(\ref{eq:sigma_c}), which for the microscopic definitions chosen above can be expressed for any initial state in the form of Eq.~(\ref{eq:ClausiusEntropy}). It is not hard to confirm that this can be decomposed as 
\begin{equation}\label{eq:Clausius_pure}
 \begin{split}
  \Sigma(t) =&~ I_{S-B}(t) + \sum_\nu I_{\circ\nu}(t) + I_{L-R}(t) \\ 
  &+ D_\text{env}^*(t) - D_\text{env}^*(0),
 \end{split}
\end{equation}
where we used that for the current class of initial states $I_{S-B}(0) = I_{\circ\nu}(0) = I_{L-R}(0) = 0$. Note that the positivity of Eq.~(\ref{eq:Clausius_pure}) is no longer guaranteed due to the last term. 
Nevertheless, the example in Fig.~\ref{fig:purestateb} shows that the entropy production is positive and almost identical to the previous ensemble averaged entropy production. This has been confirmed for various realizations of the initial state (not shown here for brevity). Typicality arguments then suggest that a negative entropy production is a rare exception for the here considered class of initial states and dynamics.

Furthermore, Fig.~\ref{fig:purestateb} (a-4) and (b-4) show that the final correlations $I_{S-B}(t)$, $I_{\circ\nu}(t)$ and $I_{L-R}(t)$ are larger for pure initial states compared to the ensemble case, which appears reasonable upon recalling that pure quantum states can have stronger correlations than mixed states. However, since the value of $\Sigma(t)$ [and also of $\sigma(t)$] agree very well in both cases, this implies that the stronger contributions from the correlations are compensated by the substraction of the term $D_\text{env}^*(0)$, which quantifies how much each mode differs initially from the corresponding grand canonical ensemble. Indeed, for our choice of pure initial states this term is certainly non-zero, which is explicitly shown in Fig.~\ref{fig:purestateb} (a-5) and (b-5). Indeed, we see that the last line in Eq.~(\ref{eq:Clausius_pure}) adds up to a small negative contribution (discarding fluctuations) in stark contrast to the ensemble case.

It is worth to think about how our results would change for non-integrable systems. Indeed, from the eigenstate thermalization hypothesis~\cite{DAlessioEtAlAP2016, DeutschRPP2018} one would naturally expect that in regime III small patches of the baths should look locally thermal in unison with the (effective) global thermal ensemble. This would imply $D_\text{env}^*(t)\approx0$ in regime III, but we observe a strong contribution from the local athermality, see Fig.~\ref{fig:purestateb} (a-5) and (b-5). This implies that the microscopic bath state is \emph{not} sampling typical (``Haar random'') states~\cite{FarquharLandsbergPRSLA1957, BocchieriLoingerPR1958, GemmerMichelMahlerBook2004, GoldsteinEtAlPRL2006, PopescuShortWinterNatPhys2006}. This might be a distinctive feature of non-integrable systems, as also studied in Ref.~\cite{BianchiEtAlPRXQ2022}. Thus, it seems as if different microscopic contributions to the entropy production scale differently in integrable and non-integrable system, albeit adding up to the same final value. Care is required with this assertion as our conclusions rely on representing the bath as a set of non-interacting Fermions, something which is not possible for non-integrable systems and even in the integrable case different geometries of the bath (e.g., a chain representation~\cite{WoodsEtAlJMP2014, StrasbergEtAlPRB2018, SchallerNazirBook2018}) can give rise to a different microscopic decomposition of the entropy production. Nevertheless, it would be interesting to look for fundamental differences of these microscopic components between integrable and non-integrable system in the future.

\section{Conclusion}\label{sec:conclusions}

Based on exact numerical integration of the Schr\"odinger equation for a paradigmatic quantum transport model, we have studied the dynamics of entropy production, its information-theoretic components, as well as temperature and chemical potential across all relevant time scales (excluding late time Poincar\'e recurrence). This was done within a precise microscopic framework and within both the conventional ensemble approach and a pure state approach. 

We highlight the following seven (to us non-trivial) insights obtained in this paper. (1) For transient and late times the entropy production is dominated by a competition between the build-up of correlations and microscopic deviations from thermal equilibrium (athermality). (2) Despite the microscopic athermality, we observed thermalization of the relevant macroscopic observables in an \emph{integrable} system for \emph{both} pure and mixed states. (3) Going beyond late time thermalization, the behaviour of the macroscopic observables also agreed at earlier times for pure and mixed states. (4) In addition to macroscopic observables, also abstract information theoretic quantities equilibrate to the same value in both baths. (5) The effective nonequilibrium temperatures and chemical potentials of Ref.~\cite{strasberg2021first, strasberg2021clausius, quantum2022Strasberg} work very well across all time scales and in unison with thermodynamic expectations. (6) We also observed that the interbath correlations are fairly insensitive to the coupling strength, which implies that the real state is far from a decorrelated state regardless of thermalization observed from a macroscopic point of view.
(7) Despite being small in an absolute sense, the system-bath correlation dominates by far all other correlations in a relative sense. 

Our results raise several questions and open up much room for future investigations. First, the single electron transistor is a very simple and effective model, which should certainly not be taken as a literal description of the physical system-bath model. Whether the seven points above hold for realistic models is thus an open question. However, at least our conclusions seem quite robust for this model: we have observed them for a wide range of coupling strengths, initial temperatures and chemical potentials, for symmetric and asymmetric baths, and for both pure and mixed states. Moreover, the observation of consistent thermodynamic behaviour across all time scales for such a simple and in particular integrable model raises our confidence in the general applicability of our results. 

Among the foundational questions that arise from this research we emphasize the point~(2) that macroscopic thermalization is observed despite strong deviation from Gibbs states and highlight that the widely held believe that non-integrability and chaos is essential for the foundations of statistical mechanics requires further scrutiny. In this respect, further studies are also required to understand our results in relation to von Neumann's $H$-theorem~\cite{beweis1929Neumann, proof1929Neumann} and recent studies of more general coarse-grained (or observational) entropies~\cite{quantum2019Safranek, strasberg2021first, brief2021Safranek, quantum2022Strasberg, nonconjugate2022Stokes}. 


From an applied perspective, our model opens up the possibility to understand cold atoms experiments~\cite{BrantutEtAlScience2012, BrantutEtAlScience2013, HaeuslerEtAlPRX2021} from first principles, without relying on any ensemble averages. 
Moreover, it is interesting to benchmark our results with master equations that have been specifically developed to deal with dynamically changing environments~\cite{EspositoGaspardPRE2003a, BreuerGemmerMichelPRE2006, EspositoGaspardPRE2007, RieraCampenySanperaStrasbergPRXQ2021, RieraCampenySanperaStrasbergPRE2022}. Finally, we believe that the interplay between the system-bath coupling strength and system-bath correlations, and their impact on the dynamics of open quantum systems, requires further investigation in light of the widely used Born approximation~\cite{nonadditive2018mitchison, KolovskyPRE2020, StrasbergEtAlPRA2023, VanKampenPhys1954}. 


\section{Acknowledgements}

AU is financially supported by Maria de Maeztu project (Grant CEX2019-000918-M funded by MCIN/AEI/10.13039/501100011033) and acknowledges further support from the Agencia Estatal de Investigaci\'{o}n and the Ministerio de Ciencia e Innovaci\'{o}n. 
KP is supported by the National Science Centre, Poland, under Grant No.~2017/27/N/ST3/01604, and by the Scholarships of Minister of Science and Higher Education. KP and ME are also supported by the FQXi foundation project ``Colloids and superconducting quantum circuits'' (Grant No. FQXi-IAF19-05-52).
PS is financially supported by ``la Caixa'' Foundation (ID 100010434, fellowship code LCF/BQ/PR21/11840014) and acknowledges further support from the European Commission QuantERA grant ExTRaQT (Spanish MICINN project PCI2022-132965), by the Spanish MINECO (project PID2019-107609GB-I00) with the support of FEDER funds, the Generalitat de Catalunya (project 2017-SGR-1127) and by the Spanish MCIN with funding from European Union NextGenerationEU
(PRTR-C17.I1).

\appendix

\begin{figure*}[tb]
\centering
\includegraphics[width=.95\linewidth]{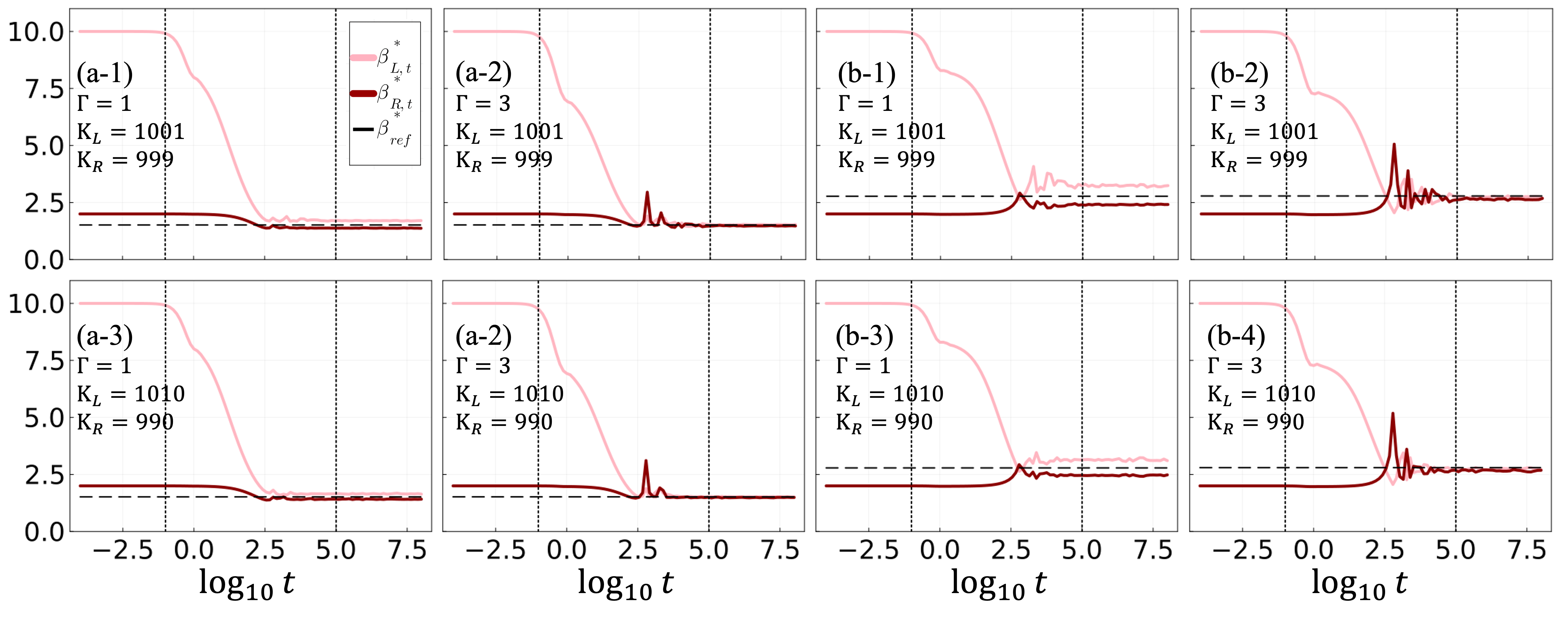}
\caption{
Effective inverse temperatures $\beta_{L,t}^*$, $\beta_{R,t}^*$ when changing the coupling strength $\Gamma$ in the case of asymmetric baths $K_L\neq K_R$. The first and second rows correspond to the case where $K_L=1001$ and $K_R=999$ and the case where $K_L=1010$ and $K_R=990$, respectively. The dashed lines are the temperatures of the global Gibbs ensemble. 
}
\label{fig:efftemchemi_asy}
\end{figure*}

\section{Asymmetric baths}\label{app:asymmetry}

In the main text, we assumed the same number of modes $K$ in each bath, which implies $\epsilon_{Lk} = \epsilon_{Rk}$ for all $k$. Here, we break the symmetry of the baths by considering a different number of modes $K_L\neq K_R$. Clearly, if our claim about the universality of our results for this model is correct, an asymmetric bath structure should not change our conclusions. 

The results are displayed in Fig.~\ref{fig:efftemchemi_asy}, where, for brevity, we only study the evolution of inverse temperature. In the first row we consider baths with the slightest possible asymmetry of $K_L=1001$ and $K_R=999$ for case (a) and (b) as studied in the main text. We do this for two coupling strength: $\Gamma = 3$ as also considered in the main text and an even weaker coupling strength $\Gamma = 1$. We see that slight discrepancies in regime III start to develop for $\Gamma=1$. This is due to the fact that in a finite sized system the coupling strength cannot be arbitrarily weak, but has to be above a certain threshold to make the states in the bath interact sufficiently well with the system state. The first row of Fig.~\ref{fig:efftemchemi_asy} precisely indicates that this threshold is around $\Gamma\approx3$.

In the second row of Fig.~\ref{fig:efftemchemi_asy}, we set $K_L=1010$ and $K_R=990$, thereby increasing the ``asymmetry parameter" $|K_L-K_R|/(K_L+K_R) = 0.01$ by a factor 10 compared to the first row. Still, the inverse temperatures evolve in the same way as above because each bath is large enough and the level spacing $\Delta\epsilon\approx W/K_\nu$ is small enough. In fact, we have confirmed thermalization up to an asymmetry of $K_L=1500$ and $K_R=500$ provided the coupling strength was not chosen unreasonably small (results not shown here for brevity). 

Finally, we remark that we refrained from showing explicit results for the chemical potential or for the entropy production and its components because we observe qualitatively the same behaviour as in the main text.

\section{Global grand canonical ensemble}\label{app:globalgibbs}

To compare the prediction of the local bath temperatures and chemical potentials with the global equilibrium state, we construct the global grand canonical ensemble according to the fixed (because conserved) global energy and particle number. This is done as follows.

First, we diagonalize the total Hamiltonian $\hat{H}$ using a new set of fermions $\{\hat{f}_n^{\dagger},\hat{f}_n\}$ such that $\hat{H}=\sum_{i,j=1}^{2K+1}\mathcal{H}_{ij}\hat{c}_i^{\dagger}\hat{c}_j=\sum_{n=1}^{2K+1}\tilde{\mathcal{H}}_{nn}\hat{f}_n^{\dagger}\hat{f}_n$ with $\tilde{\mathcal{H}}$ the diagonal matrix. Note that we relabelled $d$ as $1$, $Lk$ as $k+1$, and $Rk$ as $K+1+k$. The eigenvectors $\psi_n(j)$ of $\mathcal{H}$ connect $\hat{f}_n$ and $\hat{c}_j$ as
\begin{align}
    \hat{f}_n
    &=
    \sum_{j=1}^{2K+1} \psi_n(j) \hat{c}_j
    .
\end{align}
Thus, the global grand canonical ensemble in the eigenbasis is given by $\pi(\beta,\mu) = \mathrm{e}^{-\sum_{n=1}^{2K+1}\beta(\tilde{\mathcal{H}}_{nn}-\mu)\hat{f}_n^{\dagger}\hat{f}_n}/Z$ with the (global) temperature $\beta$ and the chemical potential $\mu$. Furthermore, its correlation matrix $\mbox{tr}[\hat{f}_n^{\dagger}\hat{f}_m \pi(\beta,\mu)]$ in the eigenbasis is a diagonal matrix with the Fermi distribution $f(\beta (\tilde{\mathcal{H}}_{nn}-\mu))$. By using the coefficients $\psi_{n}(j)$, we can rewrite this correlation matrix $\mbox{tr}[\hat{f}_n^{\dagger}\hat{f}_m \pi(\beta,\mu)]$ in the original basis as
\begin{align}
    \mbox{tr}[\hat{c}_i^{\dagger}\hat{c}_j \pi(\beta,\mu)]
    &=
    \sum_{n=1}^{2K+1} 
    \psi_n(i) \psi_n^{*}(j) 
    \mbox{tr}[\hat{f}_n^{\dagger}\hat{f}_n \pi(\beta,\mu)]
    .
\end{align}
Then, we obtain the inverse temperature $\beta_{\text{ref}}$ and chemical potential $\mu_{\text{ref}}$ by adapting the defining equations (\ref{eq:betaeff}) and (\ref{eq:mueff}) to the global grand canonical ensemble by demanding that its energy and particle number should match the true and conserved expectation value of energy and particle number.

\bibliography{bibliografia}

\end{document}